\documentstyle[multicol,prb,aps,epsf]{revtex}
\begin{document}
\title{Phonon line shape in disordered A$_3$C$_{60}$ (A= K, Rb)}

\author{J.E. Han$^{(1)}$, O. Gunnarsson$^{(1)}$ and V. Eyert$^{(1,2)}$} 
\address{${}^{(1)}$Max-Planck-Institut f\"ur Festk\"orperforschung, 
D-70506 Stuttgart, Germany}
\address{${}^{(2)}$Institut f\"ur Physik, Universit\"at Augsburg,   
D-86135 Augsburg, Germany}

\date{\today}
\maketitle
\begin{abstract}
We present a calculation of the H$_g$ phonon self-energy for a 
model of A$_3$C$_{60}$ (A= K, Rb). The orientational disorder of 
the C$_{60}$ molecules is included, and the lowest order self-energy
diagram is considered.
The calculations illustrate that due to the violation of momentum
conservation by the orientational disorder, Raman scattering can 
measure the decay of a phonon in an electron-hole pair, allowing 
the estimate of the electron-phonon coupling from such experiments. 
Comparison with experimental line shapes further provides support 
for a local correlation of the molecular orientations, with 
neighboring molecules preferentially having orientations differing
by a 90$^{\circ}$ rotation. 
\end{abstract}
\begin{multicols}{2}
\section{Introduction}
It is essential for the understanding of the alkali-doped
fullerenes A$_3$C$_{60}$ (A= K, Rb) to obtain reliable 
estimates of the strength $\lambda$
of the electron-phonon interaction. In metallic systems, such as
A$_3$C$_{60}$, a phonon can decay in an
electron-hole pair. This decay leads to an extra width of the phonon.
It was pointed out by Allen\cite{Allen} that $\lambda$ can be 
estimated from the average of this extra width ($\gamma_{ph}$)  
over all ${\bf q}$-vectors. This width can be measured in, e.g., 
neutron scattering.\cite{Prassides} It is, however, not easy to 
extract accurate values of $\lambda$ from such experiments for C$_{60}$.   
Raman scattering provides an interesting alternative, since it gives 
a high resolution and since the Raman active modes (A$_g$ and H$_g$)
are the modes 
which couple to the electrons in the partly filled $t_{1u}$ band.

In the context of the fullerenes, it was early pointed out that 
for the ordered system energy and momentum  conservation forbids 
the decay of a ${\bf q}=0$ intramolecular phonon in an
intraband electron-hole pair excitation.\cite{Chakravarty}
Due to the long wave length of the photons used in Raman
scattering, this experiment would then not be appropriate
for extracting $\lambda$, since ${\bf q}\approx 0$ phonons
are excited. A$_3$C$_{60}$ has, however, strong 
orientational disorder,  with each molecules taking essentially 
randomly one out of two preferred directions.\cite{Stephens}
Schl\"uter {\it et al.} therefore emphasized  that decay of 
a H$_g$ phonon is nevertheless possible, since ${\bf q}$-conservation
is violated,\cite{Schluter} while the decay of a ${\bf q}=0$ A$_g$
phonon is not possible.\cite{Gelfand}
 It was later assumed that the violation
of ${\bf q}$-conservation is so efficient, that Raman scattering
actually measures an average of H$_g$ phonons at all 
${\bf q}$-vectors,\cite{Rice} as assumed in the Allen formula.
Recently, however, the applicability of Allen's
formula was anew put into question.\cite{Kabanov}  
It was concluded that for a ${\bf q-}$independent interaction,
there is no broadening of the ${\bf q}=0$ H$_g$ mode, and that the 
broadening of this mode is entirely due to the ${\bf q}$-dependence
of the coupling. This led to a drastic reinterpretation of the
coupling constants derived from earlier Raman measurements.\cite{Winter} 
Explicit calculations have been performed for orientationally disordered
systems, but the broadening of the ${\bf q}=0$ modes was not 
studied.\cite{Deshpande} Therefore, 
although Allen's formula is the basis for deductions of $\lambda$
from Raman data, its validity for disordered A$_3$C$_{60}$
has to our knowledge never been explicitly tested. 
The purpose of this paper is to perform such a test.

In Sec. II we present our model and the formalism, the results are
presented and discussed in Sec. III and we give a summary in Sec. IV. 
 
\section{Model and formalism}

We consider a model which includes the partly occupied $t_{1u}$ band.  
There are three $t_{1u}$ orbitals on each C$_{60}$ molecule $i$, which are 
connected by hopping matrix elements $t$
\begin{equation}\label{eq:4b}
H^{\rm el}=\sum_{i\sigma}\sum_{m=1}^3\varepsilon_{t_{1u}}n_{im\sigma}+
\sum_{<ij>\sigma
mm'}
t_{ijmm'}\psi^{\dagger}_{im\sigma} \psi_ {jm'\sigma}
\end{equation}
The orientational disorder\cite{Stephens} has been built into the
matrix elements $t_{ijmm'}$.\cite{Orientation,Satpathy,MazinAF}
We want to describe the coupling to the intramolecular five-fold 
degenerate $H_g$ Jahn-Teller modes. 
To describe the electron-phonon interaction, we use the Hamiltonian
\begin{eqnarray}\label{eq:4d}
&&H^{\rm el-ph}=
 \omega_{ph}\sum_{i,\mu=1}^5(b^{\dagger}_{i\mu}b_{i\mu}+{1\over 2})\nonumber \\ 
&&+ {g\over 2} \sum_{\mu=1}^5 \sum_{i\sigma}\sum_{m=1}^3
\sum_{m^{'}=1}^3
V_{mm^{'}}^{(\mu)}\psi^{\dagger}_{im\sigma} \psi_{im^{'}\sigma}(b_{i\mu}+ 
b_{i\mu}^{\dagger}),
\end{eqnarray}
where $\omega_{ph}$ is the a phonon frequency, $b_{i\mu}$ annihilates
a phonon with quantum number $\mu$ on site $i$, $g$ is an overall 
coupling strength and $V^{(\mu)}_{mm^{'}}$ are dimensionless coupling 
constants\cite{Lannoo,c60jt} given by symmetry;  
$$
V^{(\mu)}=\left( \begin{array}{ccc} 
	1 & 0 & 0 \\
	0 & 1 & 0 \\
	0 & 0 & -2 
	\end{array} \right),
\sqrt{3}\left( \begin{array}{ccc} 
	1 & 0 & 0 \\
	0 & -1 & 0 \\
	0 & 0 & 0 
	\end{array} \right),
$$
\begin{equation}
\sqrt{3}\left( \begin{array}{ccc}
        0 & 1 & 0 \\
        1 & 0 & 0 \\
        0 & 0 & 0
        \end{array} \right),
\sqrt{3}\left( \begin{array}{ccc}
        0 & 0 & 0 \\
        0 & 0 & 1 \\
        0 & 1 & 0
        \end{array} \right),
\sqrt{3}\left( \begin{array}{ccc}
        0 & 0 & 1 \\
        0 & 0 & 0 \\
        1 & 0 & 0
        \end{array} \right),\label{eq:v}
\end{equation}
for $\mu=1,..,5$ respectively.
The electron-phonon coupling constant $\lambda$ is then given 
by\cite{Lannoo,c60jt}
\begin{equation}\label{eq:4c}
\lambda={5\over 3}N(0){g^2\over \omega_{ph}},
\end{equation}
where $N(0)$ is the density of states (DOS) per spin at the Fermi energy.
We have assumed that the H$_g$ modes are local Einstein modes with 
a local coupling to the electrons. Since these modes are 
intramolecular and since the interaction between the C$_{60}$ 
molecules is very weak, these assumptions should be very good.
Actually, explicit calculations of the phonon dispersion
curves for the H$_g$ modes find that the dispersion is
extremely small.\cite{phonondisp}

To describe the broadening of the phonon, we calculate the phonon 
self-energy due to the electron-phonon interaction.
We consider the lowest order self-energy, expressed in terms of 
the zeroth order electron and phonon Green's functions. 
This is the formalism used by Allen.~\cite{Allen}
In the present case, the phonon energy is not much smaller 
than the bandwidth and the validity of Migdal's theorem is 
therefore questionable. Contributions beyond the lowest order 
self-energy diagram may therefore be important. The on-site
Coulomb interaction is large in these systems, which are believed to
be close to a Mott transition.~\cite{RMP}
For simplicity, we have here nevertheless neglected corrections
to Migdal's theorem and effects of the Coulomb interaction.

We diagonalize the electronic part $H^{el}$ of the 
Hamiltonian. The eigenstates are labelled by $n$, the eigenvalues
are $\varepsilon_n$  and the corresponding
zeroth order electron Green's function is $G^0_n(\omega)$. The
phonon self-energy is then
\begin{equation}\label{eq:5}
\Pi_{\alpha,\alpha^{'}}(\omega)=-i\sum_{\sigma}\sum_{nm}\int 
{d\omega^{'}\over 2\pi}
g_{nm}^{\alpha}g_{nm}^{\alpha^{'}}G^0_n(\omega^{'})G^0_m(\omega+\omega^{'}),
\end{equation}
where $\alpha\equiv (i\mu)$ labels a combined site and phonon degeneracy
index and $g_{nm}^{\alpha}$ is the matrix element of the coupling
to the phonon $\alpha$ between the one-particle states $n$ and $m$.
The frequency integral can be performed analytically, giving
\begin{equation}\label{eq:5a}
\Pi_{\alpha,\alpha^{'}}(\omega)=2\sum_{nm}
g_{nm}^{\alpha}g_{nm}^{\alpha^{'}}{f_m-f_n\over 
\varepsilon_m-\varepsilon_n-\omega+i\eta},
\end{equation}
where $f_n\equiv f(\varepsilon_n)$ is the Fermi function, $\eta$
an infinitesimal positive number.
In the present calculation, we used $\eta=0.0004$ eV, comparable to
experimental resolution~\cite{Winter}
 and much smaller than $\omega_{ph}$ and $\gamma_{ph}$.

The phonon Green's function $D$ is given by
\begin{equation}\label{eq:6}
D^{-1}=\lbrack D^0\rbrack ^{-1}-\Pi,
\end{equation}
where a matrix notation is understood and the noninteracting 
phonon Green's function is given by
\begin{equation}\label{eq:7}
D^0_{\alpha\alpha^{'}}(\omega)={2\omega_{ph}\over 
\omega^2-\omega_{ph}^2+i\eta}\delta_{\alpha\alpha^{'}}.
\end{equation}
We introduce the reduced Green's function in momentum space
\begin{equation}\label{eq:8}
\tilde D_{\mu,\mu^{'}}({\bf q},\omega)={1\over N}\sum_{ij}\sum_{\nu}
e^{i{\bf q}\cdot({\bf R_i-R_j})}c_{\mu\nu}^{ij}
D_{i\nu,j\mu^{'}}(\omega),
\end{equation}
where $N$ is the number of molecules in the cluster, $c_{\mu,\nu}^{ij}$
the scattering coefficients with proper symmetry factors corresponding to
the specific experiment, as discussed in the Appendix.

We compute the coefficients $c_{\mu,\nu}^{ij}$ for Raman scattering
experiments. We consider electronic transitions within the $t_{1u}$ and 
$t_{1g}$ bands, assuming that the photon energy is larger than the 
$t_{1u}$-$t_{1g}$ splitting. To obtain a simple result, we make the 
approximation that the hopping matrix elements are small. Calculations 
using realistic hoppings give qualitatively similar results.  
Details of the derivation are given in the Appendix. Then, we have 
an expression for $c_{\mu\nu}^{ij}$ which only depends on $\mu,\nu$ 
and molecular orientations $o(i)$ at sites $i,\ j$, $(o(i)=\pm 1)$.

Between molecules with the same orientation, {\it i.e.,}
$o(i)=o(j)$,
\begin{equation}
c_{\mu,\mu}^{ij}=1\ (\mbox{all $\mu$}),\label{raman:eq1}.
\end{equation}
For $o(i)=1,o(j)=-1$,
\begin{equation}
c_{11}^{ij}=-c_{22}^{ij}=-c_{33}^{ij}=-c_{45}^{ij}=c_{54}^{ij}=1,
\end{equation}
and for $o(i)=-1,o(j)=1$,
\begin{equation}
c_{11}^{ij}=-c_{22}^{ij}=-c_{33}^{ij}=c_{45}^{ij}=-c_{54}^{ij}=1.
\end{equation}
Finally,
\begin{equation}
 c_{\mu\mu}^{ij}=0, \ \mbox{otherwise}.\label{raman:eq3}
\end{equation}

In the following we are interested in $\tilde D_{\mu,\mu}({\bf q}
=0,\omega)$, since in Raman scattering the coupling is to 
$\sum_{\mu}\tilde D_{\mu,\mu} ({\bf q}=0,\omega)$,
and  in
\begin{equation}\label{eq:9}
D^{\rm local}(\omega)={1\over N}\sum_{\bf q}\sum_{\mu}
\tilde D_{\mu,\mu}({\bf q},\omega).
\end{equation}
In the ${\bf q}=0$ phonon Green's function momentum conservation
is considered, to the extent that it survives the orientational 
disorder. On the other hand, the usage of      
$D^{\rm local}$ instead implies an assumption that the 
 orientational disorder has completely destroyed momentum
conservation and that we therefore can average over all momenta
as in neutron scattering. The spectral functions are defined 
as the imaginary part of the Green's functions, {\it i.e,} 
$\rho_{\mu}({\bf q},\omega)=
-(1/\pi) {\rm Im} \tilde D_{\mu,\mu}({\bf q},\omega)$.

Allen's formula for the phonon broadening $\gamma_{ph}$  (full width
at half maximum (FWHM)), can be derived from 
Eq. (\ref{eq:5a}) in the limit of $w_{ph} \ll W$ and by ignoring the 
momentum conservation, and it is given by 
\begin{equation}
\gamma_{ph}=\frac{2\pi}{3}g^2\omega_{ph} N(0)^2=
\frac{2\pi}{5}\lambda\omega_{ph}^2 N(0).
\label{eq:allen1}
\end{equation}
The phonon level shift estimated from the real part of 
the self-energy can be expressed as
\begin{equation}
\Delta \omega_{ph}=\left(-2\ln2+{7\over 9}(w_{ph}N(0))^2\right) g^2 N(0),
\label{eq:allen2}
\end{equation}
for a flat DOS. The second term in the bracket is the correction 
of order $(\omega_{ph}/W)^2$.

\section{Results}

We have performed calculations for clusters with 256 atoms
on an fcc lattice and with periodic boundary conditions.
We take into account the fact that neighboring molecules
tend to anti-align,\cite{Teslic} due to the hopping integral 
being stronger for such configurations.\cite{Orientation,MazinAF}
We control the randomness of orientation 
by introducing an Ising-type nearest neighbor antiferromagnetic
interaction for the two preferred orientations, {\it i.e.,}
$E_{rot}={\cal N}^{-1}\sum_{<i,j>}o(i)o(j)$ with
${\cal N}$ the number of nearest neighbor pairs. 
First we start from a truly random
orientation and then {\it anneal} the system to any desired 
randomness by using a ficticious temperature. Since the fcc lattice is
frustrated, there are many ``antiferromagnetic'' (AFM) states which minimize
$E_{rot}$ at $E_{rot,min}=-1/3$. Therefore even a perfectly
annealed system retains a certain disorder,
as can be seen in the electron density
of states. Then the phonon spectral functions are averaged over many 
sets of orientational configurations.

\begin{figure}
\centerline{\epsfxsize=3.3in \epsffile{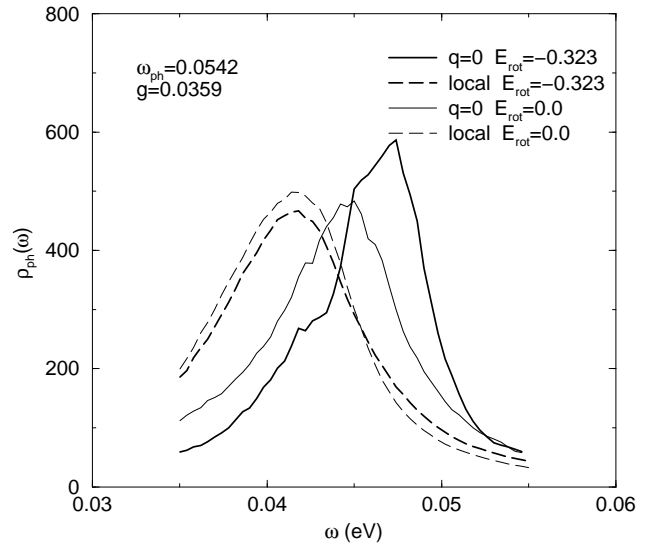}}
\caption[]{{\bf q}=0 and local phonon spectral functions at different 
rotational energies. The local spectral functions (dashed lines)
show structureless
profile with the weak dependence on the rotational energy. The $q=0$
spectral functions significantly depend on the disorder structure
of the fcc C$_{60}$ lattice.
}
\label{q0rand}
\end{figure}

The phonon spectral functions for the scattering vector {\bf q}=0 and 
the local (summed over all {\bf q}) spectral function are plotted in
Fig.~\ref{q0rand} for different orientational disorders. 
The phonon line widths for {\bf q}=0 (solid lines) and local 
(dashed lines) spectral functions are about the same 
over a large range of disorder (different values of $E_{rot}$).
This shows that the orientational disorder effectively breaks
momentum conservation, and it supports the assumption
that the Raman scattering experiment can give an estimate of the
electron-phonon coupling. On the other hand, a calculation for a
periodic ``ferro-magnetically'' ordered system gave two unbroadened
peaks due to energy and momentum conservation, as will
be discussed later.  

The validity of Allen's formula, Eq. (\ref{eq:allen1}), is explicitly
tested by changing the phonon frequency $w_{ph}$. Since the self-energy
is trivially proportional to $g^2$, we concentrated on the linear dependence
of $\gamma_{ph}$ on $w_{ph}$ for a fixed $g$.                      
The numerical estimates for the broadening $\gamma_{ph}$ and the level shift
$\Delta\omega$ are computed by fitting the {\bf q}=0 spectral
functions to a single-pole phonon spectral function
\begin{equation}
  \rho_{fit}(\omega)=-{\rm Im}\frac{A/\pi}{\left({\omega^2-\omega_{ph}^2\over 
  2\omega_{ph}}\right)^2-\Delta\omega+i\gamma_{ph}},
\end{equation}
where $A$ is the total spectral weight.~\cite{fit}
Although the numerical spectral 
functions are far from being single-pole structured,
the peak position and the FWHM, $\gamma_{ph}$,
can be understood in an average sense. 

The widths and shifts obtained from these fits are compared with the    
results of Allen's formula in Table ~\ref{tb:allen}. 
In Allen's formula we have used $N(0)=6.2 {\rm eV}^{-1}$, which 
is the appropriate value for our model of $A_3$C$_{60}$.
The coupling constant $g$ was set to 0.03 eV.
We find that Allen's formula is reasonably well reproduced from
the {\bf q}=0 spectral functions. The discrepancies between the 
numerical and Allen's estimated values range within 20-40
\%, which we attribute mainly to finite size effects. 
The phonon broadening decreased as the number of sites was increased.
Also note that $\gamma_{ph}$ shows a poorer agreement 
for small values of $\omega_{ph}$, where finite size 
effects become more pronounced due to the small number of 
particle-hole pairs with excitation energy
less than $\omega_{ph}$.  Due to computational limitations, we 
have considered a maximum number of 256 sites although the 
convergence with respect to the system size
is not good enough for a systematic finite size scaling analysis.

We have calculated the Raman scattering spectra for systems close to the
local minima of anti-ferromagnetically ordered orientations by
using realistic parameters for the coupling constant $g$ obtained from the 
photoemission spectra analysis\cite{PES}.
We then discuss the relevance of our calculation to Raman scattering 
experiments.\cite{Winter} We have performed
the calculations for the first two $H_g$-modes at frequencies of 
0.0336 and 0.0542 eV (270, 432 cm$^{-1}$, respectively) and
have taken average over 10 sets of rotational configurations
for $E_{rot}=-0.323$, close to the AFM minima at $E_{rot}=-0.333$. 

\begin{figure}
\centerline{\epsfxsize=3.3in \epsffile{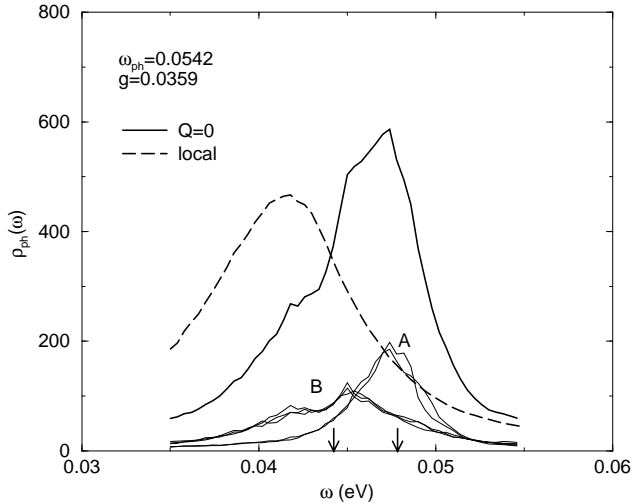}}
\caption[]{Phonon spectral functions for realistic parameter 
$w_{ph}=0.0542$ eV and $g=0.0359$ eV for $H_g(2)$ mode. The 
thick solid line shows the {\bf q}=0 phonon spectral function 
as can be observed in Raman scattering. The spectrum has 
underlying structures which can be resolved to two peaks $A$ 
and $B$ by plotting projected spectral functions for each phonon 
modes (thin lines). The local phonon spectral function (dashed 
line) shows no such underlying structure. For comparison, the peak
positions for a completely ordered periodic lattice are marked
by arrows.
}
\label{fig542}
\end{figure}

For the first two $H_g$ modes the fitted $\gamma_{ph}$'s are 13, 47 cm$^{-1}$ 
for $\omega_{ph}$=270, 432 cm$^{-1}$, respectively. These values differ
by almost a factor of 2 from the experimental estimates at 20 and 21
cm$^{-1}$, respectively.~\cite{Winter} This illustrates the 
fact that photoemission and Raman scattering give rather different      
estimates of the distribution of the coupling strength for the 
low-lying phonons.\cite{RMP} The calculated Raman spectrum 
for $H_g(2)$ mode is  shown in Fig.~\ref{fig542}. 
The total spectrum can be resolved in two major peaks
from the first two ($\mu=1,2$ at $A$) and the last three~\cite{splitB}
($\mu=3,4,5$ at $B$) modes of the phonon. This leads to a spectrum
which agrees reasonably well with the low-energy part of the spectrum 
in Ref.~\onlinecite{Winter}, in the sense that the spectrum is 
strongly skewed towards higher energies and has additional structures.
This provides support for the belief\cite{Orientation,MazinAF,Teslic}
that there is a local `antiferromagnetic' correlation, since the 
agreement with experiment is better than for the random orientation.
The two narrow high frequency peaks in the experimental spectrum seem,
however, to be beyond our treatment.

To discuss the phonon level splitting into
a doublet ($\mu=1,2$) and a triplet ($\mu=3,4,5$),
we define the reduced phonon self-energy at {\bf q}=0
in analogy to the reduced phonon Green's function
$\tilde D_{\mu,\mu'}({\bf q}=0,\omega)$ as 
\begin{equation}\label{eq:self}
\tilde \Pi_{\mu,\mu'}({\bf q}=0,\omega) = {1\over N}\sum_{ij}
\sum_{\nu}c^{ij}_{\mu\nu}\Pi_{i\nu,j\mu'}(\omega),
\end{equation}
which is the corresponding self-energy for 
$\tilde D_{\mu,\mu'}({\bf q}=0,\omega)$. (See Appendix.)
The real part of $\tilde \Pi_{\mu,\mu}(0,\omega_{ph})$ agrees very
well with the phonon level shift previously
obtained by fitting the spectral 
function for $\tilde D_{\mu,\mu}(0,\omega)$. The level shifts
are better understood by investigating the imaginary part of
$\tilde \Pi_{\mu,\mu}(0,\omega)$, which, related to the real part by the
Kramers-Kronig relation, gives the excitation spectrum responsible for 
the perturbational level shift.
Spectra of Im$\tilde \Pi_{\mu,\mu}(0,\omega)$ extend from 0 to the 
bandwidth and show different shapes for the doublet and triplet.
For orientational configurations close to the AFM minima,
those for the doublet are skewed to higher energy giving the smaller
contribution to the level shift while those for triplets are
skewed to lower energy leading to the larger level shift.
For random orientations, the two spectra show essentially the same
shape as the system size is increased.

The imaginary part of $\tilde \Pi_{\mu,\mu}(0,\omega)$
represents the spectral distribution of coupling of $\mu$-th phonon
to electron-hole pair excitations. More specifically, 
$\tilde \Pi_{\mu,\mu}(0,\omega)$ can be written as the energy spectrum
of the state for the uniformly distorted field,
$|\phi^{(\mu)}\rangle$, due to $\mu$-th Jahn-Teller phonon mode;
\begin{equation}
{\rm Im}\tilde \Pi_{\mu,\mu}(0,\omega)={\pi g^2\over 4}
{\sum_{\sigma,\alpha,\beta}}^{'} \left|\langle 
\alpha,\beta|\phi^{(\mu)}\rangle\right|^2\delta(\omega-
\varepsilon_{\alpha} +\varepsilon_{\beta}),
\end{equation}
where the summation is over the electron-hole pair states 
$|\alpha,\beta\rangle$
($\varepsilon_{\alpha}>\epsilon_F$, $\varepsilon_{\beta}
<\epsilon_F$ with the Fermi energy $\epsilon_F$) and
\begin{equation}
|\phi^{(\mu)}\rangle=\sum_{imm^{'}\sigma} \tilde\psi^\dagger_{im\sigma}
V^{(\mu)}_{mm^{'}}\tilde\psi_{im^{'}\sigma}|0\rangle,
\end{equation}
with the non-interacting Fermi sea $|0\rangle$. $\tilde\psi_{im\sigma}$
is the annihilation operator for the orbital pointing along the $m$-th principal
axis, {\it not} along the molecular orientation, {\it i.e.,}
$\tilde\psi_{im\sigma}=\psi_{im\sigma}$ for $o(i)=1$ and 
$\tilde\psi_{ix\sigma}=\psi_{iy\sigma}, \tilde\psi_{iy\sigma}=
-\psi_{ix\sigma}$ and $\tilde\psi_{iz\sigma}=\psi_{iz\sigma}$
for $o(i)=-1$.
The energy expectation value of the distorted field can be related to
the first moment of Im$\tilde \Pi_{\mu,\mu}(0,\omega)$ as
\begin{equation}
{\langle\phi^{(\mu)}|H^{el}|\phi^{(\mu)}\rangle\over
\langle\phi^{(\mu)}|\phi^{(\mu)}\rangle}=E_0+
{\int_0^\infty \omega {\rm Im}\tilde \Pi_{\mu,\mu}(0,\omega)d\omega\over
\int_0^\infty {\rm Im}\tilde \Pi_{\mu,\mu}(0,\omega)d\omega},
\end{equation}
where $E_0$ is the ground state energy of non-interacting electrons.
As expected near the AFM local minima, 
$\langle\phi^{(\mu)}|H^{el}|\phi^{(\mu)}\rangle /
\langle\phi^{(\mu)}|\phi^{(\mu)}\rangle$ has smaller value for 
the doublet and larger value for the triplet. 

The Jahn-Teller phonon couples to the distortion of the electronic
density. For the doublet modes, the phonons couple to the density
distortion along the principal axes, {\it e.g.,}
for $\mu=1$, $(b_{i1}^\dagger+
b_{i1})$ couples to $(n_{ix}+n_{iy}-2n_{iz})$, the distortion 
elongated or squeezed along $z$-axis. Similarly, the triplet phonons
couple to distortions along diagonal directions in $xy, yz, zx$-planes
{\it e.g,} 
for $\mu=3$, $(b_{i3}^\dagger+b_{i3})$ couples to $(n_{i\tilde x}-
n_{i\tilde y})$ where $\tilde x,\tilde y$ axes are rotated from $x,y$-axes
by 45$^\circ$ along $z$-axis. 
Therefore, the distortion for
$\mu=3,4,5$ tend to distribute more electrons along the in-plane
diagonal directions, which are the strongest bonding
directions for $t_{1u}$-orbitals in the fcc lattice.~\cite{Satpathy}
Furthermore, upon the inter-site coherence
close to AFM minima, the distortions from neighbors add up constructively 
along the strongest bonding direction
to give the energy gain for the triplet phonons, leading to
the skewedness toward the low energy.

It has been suggested~\cite{Winter} that the locally broken cubic
symmetry, due to the orientational disorder, could lead to a 
splitting of all five levels in the $H_g$ modes.  However within 
our model, it seems unlikely that the local energy splittings
survive after averaging over all sites. Rather, the calculation indicates
that the local energy splittings contribute to the broadening of
levels, not to collective {\bf q}=0 modes which survive disorder. We observe 
that the first two peaks in the experiment of Winter {\it et al.}
could reflect states in isolated C$_{60}$, since their observed energy 
shifts from the undoped C$_{60}$ values are too small compared to 
estimates of the phonon self-energy from the lowest order diagram
and sometimes even have different signs. 
Considering the narrowness of these peaks,
they could result from a many-electron molecular states, 
due to effects neglected an this study, and with a very weak 
coupling to the surrounding.

Our results are in strong contrast to the results of
Aksenov and Kabanov.\cite{Kabanov} They  studied the problem
in a model where the orientational disorder was described
by a relaxation rate, leading to an imaginary self-energy in 
the electron Green's function. However, the breaking of the 
translational symmetry was not explicitly included, and 
momentum conservation was assumed in evaluating various diagrams.
The orbital degeneracy was also neglected. It was concluded
that for a ${\bf q}$-independent electron-phonon coupling 
there is no broadening of a H$_g$ phonon and that the broadening 
is entirely due to the dependence of the coupling on
the angle of ${\bf q}$. This is in strong contrast to the results
above, where we find that a ${\bf q}$-independent coupling gives
a substantial broadening, provided that the violation of the 
translational symmetry is included explicitly. As was discussed 
in Sec. II, the ${\bf q}$-dependence of the coupling should be 
very weak, and the effect discussed in Ref. \onlinecite{Kabanov} 
not very important.

Aksenov and Kabanov\cite{Kabanov} were not able to obtain
the angular dependence of the coupling constants and therefore 
they could not deduce the absolute magnitude of the coupling 
constants from Raman experiments, but instead estimated  
the relative coupling constants from Raman experiments.  
To obtain absolute estimates, they used coupling constants
deduced\cite{Alexandrov} from photoemission\cite{PES} for a 
free C$_{60}$ molecule and arbitrarily assumed that the 
coupling constant to the lowest mode is correct. Had they 
instead, equally arbitrarily, assumed that the coupling to 
the third   mode is correct, all values of $\lambda_{\nu}$ 
would have been about a factor of two larger and if they had 
assumed that the deduced\cite{Alexandrov} coupling to the 
eighth mode is correct, all couplings would have been zero.

\section{Summary}                 
We have demonstrated that because the orientational disorder breaks
the momentum conservation, Raman measurements of the 
$H_g$ modes in the $A_3$C$_{60}$ should see a broadening due to the decay  
in electron-hole pairs.  This enables Raman scattering experiments to
derive electron-phonon coupling strengths. Furthermore, the explicit 
calculation shows that the width of this {\bf q}=0 mode is not very 
different from the ${\bf q}$-averaged width, where the latter is described 
by Allen's formula. We further find that the local ''anti-ferromagnetic''
correlation of the molecular orientation is important for obtaining
the non-symmetric line-shape seen experimentally. This provides support 
for such a local correlation actually taking place in the real 
system. This leads to a large splitting between a two-fold and a three-fold
degenerate component. Experimentally additional structures are seen 
which are not reproduced by the present calculations. It would therefore
be interesting to include corrections to Migdal's theorem beyond the
present calculations, and effects of the Coulomb interaction.

This work has been supported by the Max-Planck-Forschungspreis.

\begin{table}
\begin{tabular}{ccccc}
$w_{ph}$ & $\Delta\omega_{ph}$ & $\gamma_{ph}$ & 
$\Delta\omega_{Allen}$ & $\gamma_{Allen}$ \\ \hline
0.03 & $-0.0076$ & 0.00328 & $-0.0076$ & 0.00217\\ 
0.04 & $-0.0064$ & 0.00402 & $-0.0074$ & 0.00290 \\
0.05 & $-0.0063$ & 0.00460 & $-0.0073$ & 0.00362 \\
0.06 & $-0.0062$ & 0.00560 & $-0.0071$ & 0.00435 \\
\end{tabular}
\caption[]{Phonon level shifts and broadening from numerical 
estimates and Allen's formula. The numerical estimates are 
from fitting of {\bf q}=0 spectral function (see text) and 
the estimates from Allen's formula use
Eqs. (\ref{eq:allen1}-\ref{eq:allen2}) with $N(0)=6.2$ per eV and
$g=0.03$ eV. 
}
\label{tb:allen}
\end{table}

\section{Appendix: coefficients 
$\lowercase{c}^{\lowercase{i}\lowercase{j}}_{\mu\nu}$ for Raman scattering}

\begin{figure}
\centerline{\epsfxsize=1.8in \epsffile{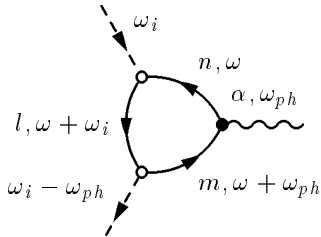}}
\caption[]{Effective photon-phonon interaction vertex. The incoming photon
(dashed lines)
with frequency $w_i$ scatters to a photon with frequency $w_i-\omega_{ph}$ by 
losing its energy to the $\alpha$-th phonon mode (wavy line) 
with energy $\omega_{ph}$. The interaction is 
mediated by the electrons (thick lines labelled as $n,m,l$) in the 
partially filled $t_{1u}$ band and in the empty $t_{1g}$ band in the solid.
The filled dot represents the electron-phonon interaction with strength
$g^\alpha_{nm}$ and the empty dots the electron-photon dipole interaction.}
\label{Raman:diag}
\end{figure}

In the Raman scattering experiment (${\bf q}=0$), the incoming light  
with energy $\omega_i$ is scattered by losing the
energy of $\omega_{ph}$ to the phonon in solid. Following
Quang {\it et al.}\cite{Quang}, we treat the photon classically. 
The scattering rate $I_p(\omega_i,\omega_{ph})$ with a polarization 
vector $p$ can then be expressed as
\begin{equation}
I_p(\omega_i,\omega_{ph})\propto\sum_{\alpha,\alpha'}
{\rm Im}D_{\alpha,\alpha'}(\omega_{ph})
h_\alpha^p(w_i,\omega_{ph})h_{\alpha'}^p(w_i,\omega_{ph})^*,
\end{equation}
where $h_\alpha^p(w_i,\omega_{ph})$ is given by, as depicted in 
Fig.~\ref{Raman:diag},
\begin{eqnarray}
h_\alpha^p(w_i,\omega_{ph})=\sum_{n,m,l}g^\alpha_{nm}P^p_{nl}P^p_{ml}
\nonumber \\   \times
\int d\omega G^0_n(\omega)G^0_m(\omega+\omega_{ph})G^0_l(\omega+\omega_i),
\label{halpha}
\end{eqnarray}
with matrix elements $P^p_{nl}$ of the dipole moment
operator at the polarization $p$ between non-interacting electronic 
eigenstates $n$ and $m$. We consider the transitions $t_{1u}\rightarrow t_{1g}$,
since these transitions have large dipole matrix elements. 
Since a typical frequency of the light source is a few eV,
and the $t_{1u}-t_{1g}$ splitting is only about $\Delta_t\sim 1$ eV,
transitions to higher states may, however, also play a role. 

We make an approximation to $h_\alpha^p(w_i,\omega_{ph})$ where the
intermolecular hoppings are ignored due to their relatively small
bandwidth. Then, the indices in the integrand in Eq. (\ref{halpha})
refer to orbitals on the same site. Since the three orbitals in $t_{1u}$
or $t_{1g}$-orbitals are all equivalent and the integral depends on
the indices only through their energies, the integral
is only a function of $w_i,\omega_{ph},\Delta_t$. 
Furthermore, since $w_i\gg W$ and
$\omega_{ph}<   W$ with the bandwidth $W\sim 0.5$ eV, we ignore the  
$w_i, \omega_{ph}$ dependency in $h_\alpha^p(w_i,\omega_{ph})$. 
Now factoring out the integral, we have
\begin{equation}
h_\alpha^p(w_i,\omega_{ph})={g\over 2}
\sum_{i,n,m,l}V^{(\mu)}_{nm}P^p_{i,nl}P^p_{i,ml},
\end{equation}
where the matrix elements $P^p_{nl}$ in Eq. (\ref{halpha}) are reduced to
matrix elements between local orbitals $n,l$ at site $i$, $P^p_{i,nl}$ 

We denote the three $t_{1u}$ and $t_{1g}$-molecular orbitals 
as $x,y,z$, as in Ref. \onlinecite{Laouini}. 
For the first type of orientation, we can easily
understand from symmetry of C$_{60}$ molecules that
\begin{equation}
P^p_{i,xy}=p_z,\ P^p_{i,yz}=p_x,\ P^p_{i,zx}=p_y\ \mbox{(orient. 1)}
\end{equation}
for a polarization direction ${\bf p}=(p_x,p_y,p_z)$ with 
$|{\bf p}|=1$. Similarly for the second type of orientation, we only need to
rotate the molecule 90 degree along z-axis, which gives
\begin{equation}
P^p_{i,xy}=p_z,\ P^p_{i,yz}=-p_y,\ P^p_{i,zx}=p_x\ \mbox{(orient. 2)}.
\end{equation}
Note that $P^p_{i,nn}=0$.
Therefore, leaving out common factors, we can write using Eq. (\ref{eq:v}),
\begin{eqnarray}
h_{i1}^p(w_i,\omega_{ph}) & = & p_x^2+p_y^2-2p_z^2,\ p_x^2+p_y^2-2p_z^2 \\
h_{i2}^p(w_i,\omega_{ph}) & = & \sqrt{3}(p_y^2-p_x^2),\ -\sqrt{3}(p_y^2-p_x^2) \\
h_{i3}^p(w_i,\omega_{ph}) & = & 2\sqrt{3}p_x p_y,\ -2\sqrt{3}p_x p_y \\
h_{i4}^p(w_i,\omega_{ph}) & = & 2\sqrt{3}p_y p_z,\ 2\sqrt{3}p_z p_x \\
h_{i5}^p(w_i,\omega_{ph}) & = & 2\sqrt{3}p_z p_x,\ -2\sqrt{3}p_y p_z,
\end{eqnarray}
where the first expressions on the right hand side are for molecules
of orientation of type 1 and the second for type 2.

Finally, we take average of $h_{i\mu}^p h_{j\mu'}^p$ over the polarization 
directions, {\it i.e.,} $<h_{i\mu}^p h_{j\mu'}^p>=\int\frac{d\Omega_p}{4\pi}
h_{i\mu}^p h_{j\mu'}^p$. Defining $c_{\mu\mu'}^{ij}=\frac{5}{4}
<h_{i\mu}^p h_{j\mu'}^p>$, we obtain $c$ as given in
 Eqs. (\ref{raman:eq1}-\ref{raman:eq3}). Then 
\begin{equation}
<I_p(\omega_i,\omega_{ph})>_p\propto \sum_{\alpha}
{\rm Im} \tilde D_{\alpha,\alpha}(\omega_{ph}),
\end{equation}
with $\tilde D_{\alpha,\alpha}$ defined in Eg. (\ref{eq:8}).

Now, we look for the Dyson equation corresponding to the reduced
Greens function $\tilde D_{\nu,\nu}({\bf q},\omega)$ at {\bf q}=0.
we can rewrite $\tilde D_{i\mu,j\nu}$ by splitting the
coefficient matrix $c^{ij}_{\mu\nu}$ as
\begin{equation}
\tilde D_{i\mu,j\nu}\equiv\sum_{\nu'}c^{ij}_{\mu\nu'}D_{i\nu',j\nu}
=\sum_{\nu',\nu''}\bar\tau^i_{\mu\nu'}D_{i\nu',j\nu''}\tau^j_{\nu''\nu},
\end{equation}
where $\tau^i_{\mu\nu}$ is the $5\times 5$ matrix defined as
the unit matrix for $o(i)=1$ and 
$$
	\left( \begin{array}{ccccc}
	1 & 0 & 0 & 0 & 0 \\
	0 & -1 & 0 & 0 & 0 \\
	0 & 0 & -1 & 0 & 0 \\
	0 & 0 & 0 & 0 & -1 \\
	0 & 0 & 0 & 1 & 0
	\end{array} \right)\mbox{ for $o(i)=-1$,}
$$ and $\bar\tau^i=(\tau^i)^{-1}$.
Now employing the matrix notation for the orbital indices,
\begin{eqnarray}
\tilde D_{ij} & = 
& \bar\tau^i\left(D^0\delta_{ij}+D^0\sum_k\Pi_{ik}D_{kj}\right)\tau^j \\
& = & D^0\delta_{ij}+D^0\sum_k\bar\tau^i\Pi_{ik}\tau^k\bar\tau^k
D_{kj}\tau^j \\
& = & D^0\delta_{ij}+D^0\sum_k\tilde\Pi_{ik}\tilde D_{kj},
\end{eqnarray}
with the reduced self-energy $\tilde\Pi_{i\mu,j\nu}(\omega)$ defined as
\begin{equation}
\tilde \Pi_{i\mu,j\nu}=\sum_{\nu'}c^{ij}_{\mu\nu'}\Pi_{i\nu',j\nu}.
\end{equation}
Finally by taking the summation over site indices $i,j$, we 
obtain the reduced phonon self-energy $\tilde \Pi_{\mu\nu}({\bf q}=0,\omega)$
for $\tilde D_{\mu\nu}({\bf q}=0,\omega)$ as Eq. (\ref{eq:self}).

\end{multicols}

\begin{thebibliography}{*}


\bibitem{Allen}P.B. Allen, Phys. Rev. B {\bf 6}, 2577 (1972);
Solid State Commun. {\bf 14}, 937 (1974).

\bibitem{Prassides}K. Prassides, C. Christides, M.J. Rosseinsky,
J. Tomkinson, D.W. Murphy, and R.C. Haddon, Europhys. Lett.
{\bf 19}, 629 (1992).


\bibitem{Chakravarty}S. Chakravarty, S. Khlebnikov, and S. Kivelson,
Phys. Rev. Lett. {\bf 69}, 212 (1992).

\bibitem{Stephens}P.W. Stephens, L. Mihaly, P.L. Lee, R.L. Whetten,
S.-M. Huang, R. Kaner, F. Diederichs, and K. Holczer, Nature {\bf 351}, 
632 (1991).
 
\bibitem{Schluter}M.A. Schl\"uter, M. Lannoo, M.F. Needels, and G.A. Baraff,
Phys. Rev. Lett. {\bf 69}, 213 (1992).

\bibitem{Gelfand}M.P. Gelfand, Supercond. Rev. {\bf 1}, 103 (1994).

\bibitem{Rice}M.J. Rice and P. Gomes da Costa, in {\it Electronic Properties 
of Novel Materials: Progress in Fullerene Research}, Eds. H. Kuzmany,
J. Fink, M. Mehring and S. Roth, World Scientific, (Singapore, 1994),
p. 501. 

\bibitem{Kabanov}V.L. Aksenov and V.V. Kabanov, Phys. Rev. B
{\bf 57}, 608 (1998).


\bibitem{Winter}   
Winter, J. and H. Kuzmany, 1996, Phys. Rev. B {\bf 53}, 655.

\bibitem{Deshpande}M.S. Deshpande, E.J. Mele, M.J. Rice, and H.-Y. Choi,
Phys. Rev. B {\bf 50}, 6993 (1994).

\bibitem{Orientation}O. Gunnarsson, S. Satpathy, O. Jepsen, and O.K. Andersen,
Phys. Rev. Lett. {\bf 67}, 3002 (1991).

\bibitem{Satpathy}
S. Satpathy,  V.P. Antropov, O.K. Andersen, O. Jepsen, O. Gunnarsson,
and A.I. Liechtenstein, Phys. Rev. B {\bf 46}, 1773 (1992).


\bibitem{MazinAF} I.I. Mazin, A.I. Liechtenstein, O. Gunnarsson, O.K.
Andersen, V.P. Antropov, and S.E. Burkov, Phys. Rev. Lett.
{\bf 26}, 4142 (1993).

\bibitem{Lannoo}
Lannoo, M., G.A. Baraff, M. Schluter, and D. Tomanek, 1991, Phys. Rev. B \
{\bf 44}, 12106.

\bibitem{c60jt}
O. Gunnarsson, Phys. Rev. B {\bf 51}, 3493 (1995).

\bibitem{phonondisp} V.R. Belosludov, and V.P. Shpakov, 
Mod. Phys. Lett. B {\bf 6}, 1209 (1992).

\bibitem{RMP} O. Gunnarsson, Rev. Mod. Phys. {\bf 69}, 575 (1997).

\bibitem{Teslic}S. Teslic, T. Egami, and J.E. Fischer, Phys. Rev.
B {\bf 51}, 5973 (1995).
 
\bibitem{fit} A simple Lorentzian fit gives significantly overestimated
values for the imaginary part of the self-energy at low frequency peaks.

\bibitem{Alexandrov}A.S. Alexandrov and V.V. Kabanov,
Pis'ma Zh. Eksp. Teor. Fiz. {\bf 62}, 920 (1995); Phys. Rev.
B {\bf 54}, 3655 (1996). In this paper the photoemission
spectrum\cite{PES} was described by including the excitation of excitons
in addition to the phonons. The exciton energy was estimated
for the initial state (C$_{60}^-$). However, the energy of the
exciton in the final state ($C_{60}$) determines if it is energetically
possible to excite an exciton at all in the photoemission process. 
For the the photon energy studied in Ref. \onlinecite{PES} this is not
the case, while studies at larger photon energies
(W. Eberhardt, priv. commun.) show an
exciton at a substantially larger binding energy than assumed
by Alexandrov and Kabanov.

\bibitem{PES}O. Gunnarsson, H. Handschuh, P.S. Bechthold,
B. Kessler, G. Gantef\"or, and W. Eberhardt, Phys. Rev. Lett.
{\bf 74}, 1875 (1995).

\bibitem{splitB} The additional structure inside the peak $B$ 
in FIG.~\ref{fig542} appears to be sensitive to the degree of the
orientational ordering and merge into one peak as the system size
becomes large. 

\bibitem{Quang} D. N. Quang, B. Esser and R. Keiper, Phys. Stat. Sol. (b)
{\bf 99}, 103 (1980).

\bibitem{Laouini} N. Laouini, O. K. Andersen, and O. Gunnarsson, Phys. Rev. B
{\bf 51}, 17446 (1995).

\end{thebibliography}
\end{document}